\documentclass[%
reprint,
superscriptaddress,
showpacs,
amsmath,
amssymb,
aps,
pra,
floatfix,
]{revtex4-1}
\usepackage{graphicx}% Include figure files
\usepackage{dcolumn}% Align table columns on decimal point
\usepackage{bm}% bold math
\begin{document}
\preprint{APS/123-QED}
\title{Optimized entropic uncertainty relation for successive measurement}
\author{Kyunghyun Baek}%
\affiliation{Department of Physics, Sogang University, Mapo-gu, Shinsu-dong, Seoul 121-742, Korea}
\author{Tristan Farrow}
\affiliation{Centre for Quantum Technologies, National University of Singapore, 3 Science Drive 2, 117543, Singapore}
\author{Wonmin Son}
\email{sonwm71@sogang.ac.kr}
\affiliation{Department of Physics, Sogang University, Mapo-gu, Shinsu-dong, Seoul 121-742, Korea}

%\affiliation{Department of Physics, Sogang University, Mapo-gu, Shinsu-dong, Seoul 121-742, Korea}
\date{\today}% It is always \today, today,
% but any date may be explicitly specified

\begin{abstract}
In the history of quantum mechanics, various types of uncertainty relationships have been introduced to accommodate different operational meanings of Heisenberg uncertainty principle. We derive an optimized entropic uncertainty relation (EUR) that quantifies an amount of quantum uncertainty in the scenario of successive measurements. The EUR characterizes the limitation in the measurability of two different quantities of a quantum state when they are measured through successive measurements. We find that the bound quantifies the information between the two measurements and imposes a condition that is consistent with the recently-derived error-disturbance relationship.
\end{abstract}


\maketitle
%\tableofcontents

\section{Introduction}
It is well-known that the Heisenberg uncertainty principle \cite{Heisenberg1927} is at the very heart of quantum mechanics. Through extensive investigations, it has been known that various distinctive properties of quantum mechanics can be derived from the principle \cite{Cohen-Tannoudji2005}. However, its precise underlying meaning has so far eluded many attempts to explain its diverse features \cite{Arthurs1965, Ozawa2001, Ozawa2004, Bush2013, Ozawa2013}. Heisenberg proposed the uncertainty relation in 1927 after postulating the kinematics of quantum canonical variables that do not commute \cite{Heisenberg1927}. It says that, as one tries to specify the position of an electron precisely, its conjugate variable, {\it e.g.} its momentum, is dispersed within a given precision. The mathematical formulation of the uncertainty was made by Kennard \cite{Kennard1927} as,
\begin{eqnarray}\label{HUR1}
\epsilon(Q)\eta(P) \geq \frac{\hbar}{2},
\end{eqnarray}
where $\epsilon(Q)$ is the mean error that occurs when an observer measures the position of an electron and $\eta(P)$ is the disturbance of the electron's momentum $P$ caused by the position measurement $Q$. $\hbar$ is the Planck constant. Relation (\ref{HUR1}) uses the statistical variances between the two measurements and was later extended to arbitrary pairs of observables by Robertson \cite{Robertson1929}. By considering generalized observables $X$ and $Y$, the lower bound is given by the commutator of the observables,
\begin{eqnarray}\label{RUR}
\delta (X)\delta (Y) \geq \frac{1}{2} |\langle\psi|[X,Y]|\psi\rangle|,
\end{eqnarray}
\noindent where $\delta (X)$ is standard deviation defined as $\delta (X)=\sqrt{|\langle\psi|(\hat{X}-\langle \hat{X}\rangle)^2|\psi\rangle|}$ and $[\hat{X},\hat{Y}]$ represents the commutator, $[X,Y]=XY-YX$. The above relation (\ref{RUR}) claims that in an arbitrary state $|\psi\rangle$, a pair of noncommuting observables cannot be well localized simultaneously.
In fact, the underlying meaning of two closely related uncertainty relations is not equivalent. Their subtle differences will become clearer when we consider the following three statements of uncertainty relations presented by P. Busch {\it et. al.} \cite{Busch2007}. Possible interpretations of the uncertainty relation can be : (i) it is impossible to {\it prepare states} in which position and momentum are simultaneously arbitrarily well localized. (ii) it is impossible to {\it measure a system's position and momentum} simultaneously. (iii) it is impossible to {\it measure position without disturbing momentum}. In these statements, position and momentum represent two conjugate variables in quantum measurement.
We can classify the above relations into three categories of physical situations. First, the Robertson relation (\ref{RUR}) is equivalent to statement (i), which identifies a fundamental limitation on preparing states whose noncommuting parameters cannot be well-localized simultaneously with arbitrary precision. This is a statement about the property of a given ensemble, not about the statistics of measured data. Second, it follows from statement (ii) that the uncertainty relations apply to the simultaneous measurement of two different variables whose measurement is impossible to implement with arbitrary precision in principle. It means that the uncertainty is the property of the statistical distributions from the measurement setup rather than the state itself \cite{Arthurs1965}. Third, Heisenberg's relation (\ref{HUR1}) is equivalent to statement (iii), since it describes the situation where a measurement of a variable, {\it e.g.} position $Q$, cannot avoid the disturbance on its conjugate variable, $P$, where $Q$ and $P$ are noncommuting observables.
Recent efforts to generalize Heisenberg's relation (\ref{HUR1}) take into account various operational circumstances by uniting statements (i), (ii), (iii). A universally valid error-disturbance uncertainty relation (EDUR) was derived in \cite{Ozawa2001} as,
\begin{eqnarray}\label{UUR}
\epsilon(X)\eta(Y)+\epsilon(X)\delta (Y)+\delta (X)\;\eta(Y) \nonumber\\ \geq \frac{1}{2} |\langle \psi| [\hat{X},\hat{Y}] |\psi\rangle|,
\end{eqnarray}
where the mean error and the disturbance are defined by $\epsilon(X)^2 =\sum_m ||M_m (m-X)|\psi\rangle||^2$ and $\eta(Y)^2 = \sum_m||[M_m,Y]|\psi\rangle||^2$, respectively, if the apparatus $M$ has a family $\{\hat{M}_m\}$ of measurement operators and $||\dots||$ denotes the norm of the state vector \cite{Ozawa2005}. This means that the measuring apparatus $M$ has possible outcomes $m$ with probability $\text{Pr}(m)=||\hat{M}_m|\psi\rangle||^2$ and the state of the object $S$ after the measurement with the outcome $m$ becomes $\hat{M}_m|\psi\rangle/||\hat{M}_m |\psi\rangle||$. It was also proved experimentally that the Heisenberg's relation (\ref{HUR1}) is violated in spin measurements, while the improved relationship (\ref{UUR}) remains valid \cite{Ozawa2012}. Later, the error-disturbance relation was improved in a stronger form \cite{Branciard2013, Bush2013}. The EDUR reduces to the Robertson uncertainty relations (\ref{RUR}) when there is no error in the first measurement $\epsilon(X)=0$ and the disturbance is replaced by the statistical deviation of the measurement $Y$ as $\eta(Y)=\delta(Y)$.
Inspired by an information theoretic interpretation of quantum uncertainty, the trade-off relation for position and momentum observables has been obtained in terms of the Shannon entropy \cite{Hirschmann1957}. The relationship was later generalized for measurements on arbitrary continuous variables \cite{Biaynicki1975}. This is called the entropic uncertainty relationship (EUR). A generalization of the EUR into the discrete observables was proposed by D. Deustch \cite{Deutsch1983} and the bound of EUR was improved by Uffink in the following form \cite{Uffink1988}. Considering observables $X$ and $Y$ with non-degenerate spectra given by $X=\sum_i x_i|x_i\rangle\langle x_i|$ and $\hat{Y}=\sum_j y_j|y_j\rangle\langle y_j|$ with the natural logarithm, the Shannon entropy $H(X)$ and $H(Y)$ is defined as
$%\begin{eqnarray}
H_\rho(X)=-\sum_i \text{Tr}[\rho |x_i\rangle\langle x_i|] \log \text{Tr}[\rho |x_i\rangle\langle x_i|]$, $ H_\rho(Y)=-\sum_j \text{Tr}[\rho |y_j\rangle\langle y_j|] \log \text{Tr}[\rho |y_j\rangle\langle y_j|]
$ %\end{eqnarray}
for a state expressed by a density matrix $\rho$. Then the EUR becomes
\begin{eqnarray}\label{EUR}
H_\rho(X)+H_\rho(Y) \geq -2\log c,
\end{eqnarray}
where the lower bound constant $c=|\max_{i,j}\langle x_i|y_j\rangle|$ is independent of the initial state. $\{|x_i\rangle\}$ and $\{|y_j\rangle\}$ are the corresponding complete sets of normalized eigenvectors with respect to operators $X$ and $Y$. In general, it can be said that EUR has a more fundamental lower bound than the variance based uncertainty relation in the sense that the bound is independent of the prepared initial state, unlike in ($\ref{RUR}$) and ($\ref{UUR}$). On the other hand, the EUR in (\ref{EUR}) is only limited by the prepared state $\rho$, like the Robertson inequality in ($\ref{RUR}$). It means that the EUR provides a fundamental constraint on the state preparation as in the case of the operational uncertainty interpretation (iii) in \cite{Busch2007}.
In this work, we have derived the uncertainty relationship characterized by the entropy under the circumstance of simultaneous measurements. We consider the case when two different measurements are performed successively on a single quantum system and find a fundamental entropic constraint which constitutes an entropic uncertainty relationship. The relationship has a different operational meaning to the original EUR in (\ref{EUR}) and is comparable to the error-disturbance versions of uncertainty relations in (\ref{HUR1}) and (\ref{UUR}). We organize our discussion as follows. In section \ref{2}, we compare the quantitative difference between the EUR and the variance based uncertainty relation. We find that they are optimized in different regimes. In section \ref{3}, the entropic uncertainty relationship for the subsequent measurements is derived and generalized. We compare each term in the relationship and discuss their optimal physical meanings. In section \ref{4}, the optimized entropic uncertainty and the error disturbance relationship are been compared and analyzed in detail. A conclusion about our results is drawn in section \ref{222}.
\section{Comparison between the EUR and the Robertson's uncertainty relation}\label{2}
\begin{figure}[t]
\begin{minipage}[h]{0.55\linewidth}
{\bf (a) $\theta=0\;\;\;\;\;\;\;\;\;\;\;\;\;\;\;\;\;\;\;\;\;$}
\end{minipage}\begin{minipage}[h]{0.25\linewidth}
{\bf (b) $\theta=3\pi/8$}
\end{minipage}
\begin{minipage}[h]{0.25\linewidth}
\includegraphics[scale =0.5]{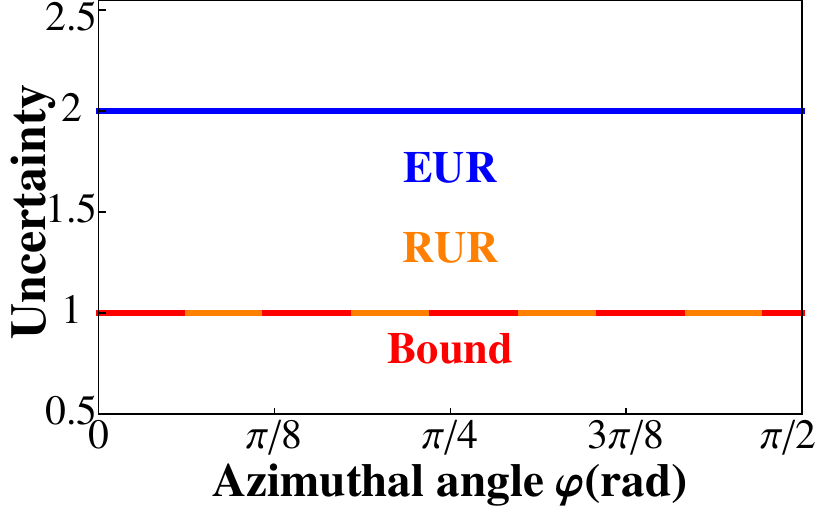}
\end{minipage}\begin{minipage}[h]{1.05\linewidth}
\includegraphics[scale =0.5]{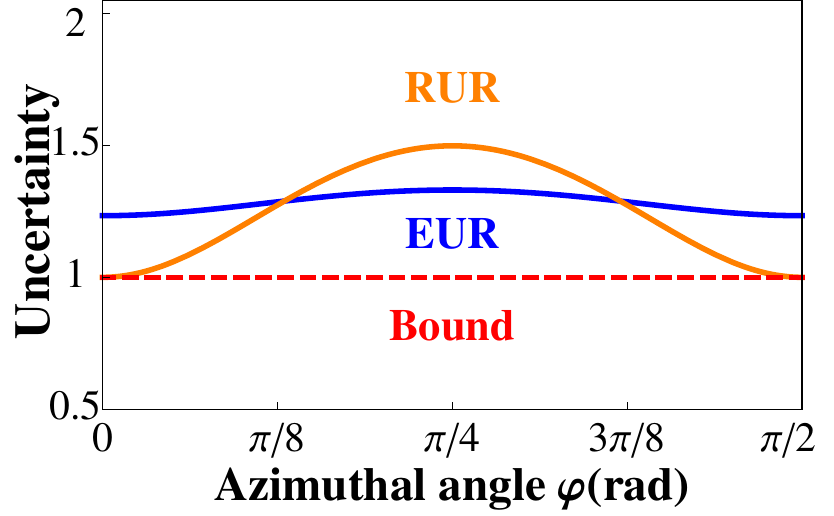}
\end{minipage}
\vspace{3 mm}
\begin{minipage}[h]{0.55\linewidth}
{\bf (c) $\theta=4\pi/9\;\;\;\;\;\;\;\;\;\;\;\;\;\;\;\;\;\;\;\;\;$}
\end{minipage}\begin{minipage}[h]{0.25\linewidth}
{\bf (d) $\theta\sim \pi/2$}
\end{minipage}
\begin{minipage}[h]{0.25\linewidth}
\includegraphics[scale =0.5]{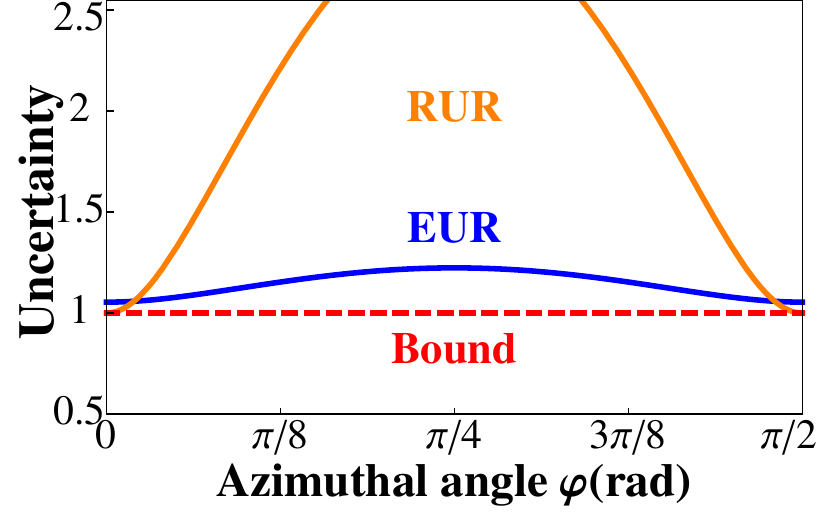}\end{minipage}\begin{minipage}[h]{1.05\linewidth}\includegraphics[scale =0.5]{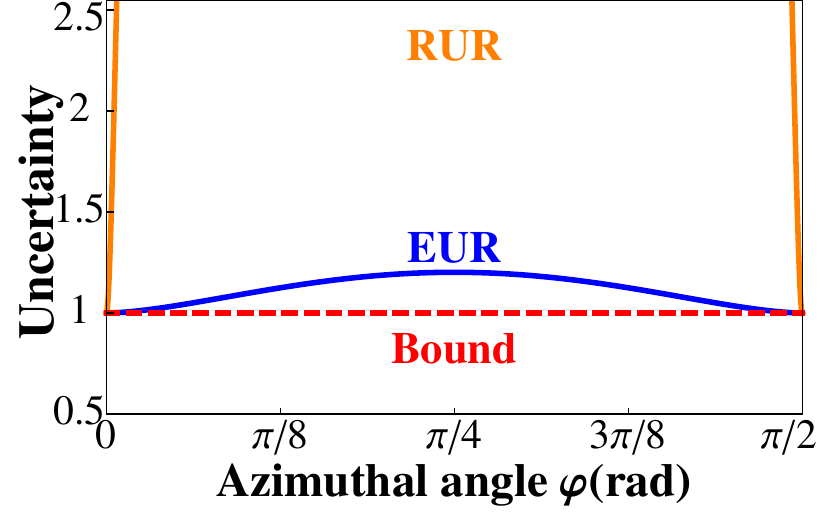}
\end{minipage}
\caption{Figures illustrating how EUR and Robertson's uncertainty relations (RUR) behave with the probabilities of outcomes of observables $X(0)$ and $Y(0)$. To compare two relations, we plot the LHS of the inequalities (\ref{DRUR}) against azimuthal angle $\varphi$ for chosen values of polar angle $\theta$ to be (a) $0$, (b) $3\pi/8$, (c) $4\pi/9$ and (d) $\pi/2$.
The EUR is optimised as $\theta$ goes to $\pi/2$ whereas RUR diverges. Relations (\ref{RUR}) and (\ref{EUR}) have a minimum value when $\theta=0$ and $\theta=\pi/2$ for fixed $\phi$, respectively.}\label{Graph-EUR}
\end{figure}

In this section, we compare the Robertson's uncertainty relation (\ref{RUR}) and the EUR (\ref{EUR}) quantitatively, to identify which is the more informative condition for a given quantum state. The former is a relationship based upon the variance of a statistical distribution and the latter is a characterization of uncertainty using Shannon's entropy.
%In the case of position and momentum observables, the entropic uncertainty relation has the form of \cite{Biaynicki1975}
%\begin{eqnarray}\label{EURCV}
%H(Q)+H(P) \geq \log \pi e \hbar
%\end{eqnarray}
%where the Shannon entropy for continuous variable is defined such that $H(Q)=-\int |\psi(q)|^2 \log |\psi(q)|^2 dq$ and $H(P)=-\int |\tilde\psi(p)|^2 \log |\tilde\psi(p)|^2 dp$ when $\psi(q)$ and $\tilde\psi(p)$ are wavefunctions of position and momentum respectively. This relation (\ref{EURCV}) is stronger than the uncertainty relation defined in terms of standard deviation. That is because the entropy of continuous variable has maximum value $\log \sqrt{2\pi e} \delta(Q)$ when a wavefunction has a fixed standard deviation $\delta(Q)$ and the maximum value is saturated when it is given by Gaussian form \cite{Shannon1948}. Hence, the satisfaction of the EUR implies the uncertainty of position and momentum observables, i.e. the Kennard's relation \cite{Kennard1927}, such as
%\begin{eqnarray}\label{EURCV}
%\log 2 \pi e \delta(Q)\delta(P)\geq H(Q)+H(P) \geq \log \pi e \hbar.
%\end{eqnarray}
A distinction between the relations is that their lower bounds behave differently: the bound for Robertson's uncertainty relation depends upon the prepared state, whereas EUR does not. It is notable that the necessity of an independent lower bound of a state has been addressed in \cite{Deutsch1983} and it is argued that such a bound is important when there is a dynamical evolution that transforms quantum states at each instance. Due to the difference, a direct comparison of the relationship is not straightforward in general.
Let us consider two general spin observables which are the simplest non-trivial example of incompatible measurements. Without loss of generality, they can be
\begin{eqnarray}
X(\phi)&=&%\sigma_\phi=
\cos\phi \;\sigma_x+\sin\phi \;\sigma_y\\
Y(\phi)&=&\sin\phi\; \sigma_x+\cos\phi\;\sigma_y
\end{eqnarray}
where $\sigma_x$ and $\sigma_y$ denote the Pauli matrices and $\phi$ characterizes the measurement angle between $X$ and $Y$. Their commutator is expressed as $[X,Y]=2i \cos2\phi~ \sigma_z$. In their two extremas, when $\phi=0$, the measurements are orthogonal and when $\phi=\pi/4$ they become identical. Once the measurement operators are specified, an analytic description of the uncertainty relations (\ref{RUR}) and (\ref{EUR}) becomes possible in general.
\begin{figure}[t]
\includegraphics[scale =0.6]{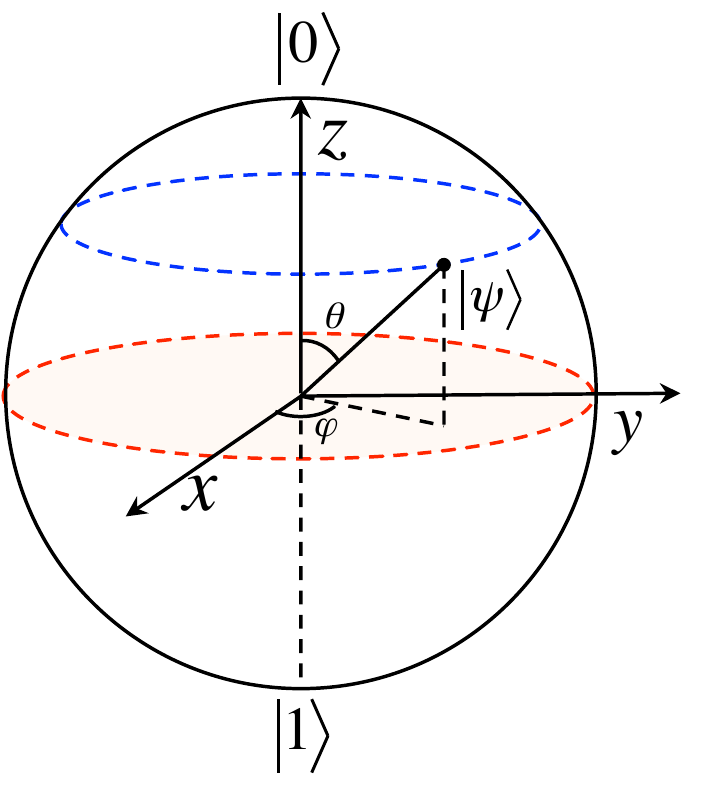}
\caption{A pure state representation in a Bloch sphere. A prepared state is denoted by $|\psi\rangle=\cos\theta|0\rangle+\sin\theta e^{i\varphi}|1\rangle$, with polar angle $0\leq\theta\leq\pi$ and azimuthal angle $0\leq\varphi\leq2\pi$. The north and south poles are chosen to correspond to eigenvectors of $\sigma_z$. Varying $\varphi$ from 0 to $\pi/2$ with a fixed $\theta$, we can consider state vectors on a circle (Dashed blue). When $\theta=\pi/2$, it becomes a circle on x-y plane (Dashed red).}\label{Bloch}
\end{figure}
A pure state is defined in a Bloch vector sphere as $|\psi\rangle=\cos(\theta/2) |0\rangle+\sin(\theta/2) e^{i\varphi}|1\rangle$ and is depicted in Fig \ref{Bloch}. In that case, the probability of outcomes for the measurements $X$ and $Y$ become
\begin{eqnarray}
p^X_{\pm}=\frac{1}{2}[1\pm \sin\theta\cos(\phi+\varphi)]\\
p^Y_{\pm}=\frac{1}{2}[1\pm \sin\theta\sin(\phi-\varphi)]
\end{eqnarray}
which can be used for the evaluation of the spin variances and entropies. For the $X$ measurement, they are
\begin{eqnarray}
\delta(X)&=&\sqrt{1-\langle \hat{X}\rangle}=\sqrt{1-(p^X_+ -p^X_-)}\\
H_{|\psi\rangle}(X)&=&-p^X_+\ln p^X_+-p^X_-\ln p^X_-
\end{eqnarray}
and similar relations can be found for the Y measurement. The entropy $H_{|\psi\rangle}(X)$ is $H_{\rho}(X)$ when a state $\rho$ is a pure state given by $\rho=|\psi\rangle\langle\psi|$. With these formulae, a direct comparison of the uncertainty relationship (\ref{RUR}) and (\ref{EUR}) can be made as follows.
The uncertainty relations can be reformulated by the normalization, meaning that both sides of the relations are divided by their own lower bound. The normalized relations have same bound $1$, such that
%\begin{eqnarray}
%\frac{H_{|\psi\rangle}(X)+H_{|\psi\rangle}(Y)}{-2\log c} &\ \geq 1,\label{DEUR}\\
%\frac{\delta(X)\delta(Y)}{|\langle [X,Y] \rangle|/2} \ \geq 1,\label{DRUR}
%\end{eqnarray}
\begin{equation}
\frac{H_{|\psi\rangle}(X)+H_{|\psi\rangle}(Y)}{-2\log c} \geq 1,
~~\frac{\delta(X)\delta(Y)}{|\langle [X,Y] \rangle|/2} \ \geq 1,
\label{DRUR}
\end{equation}
where $c=\sqrt{(1+\sin 2\phi)/2}$ for $0 < \phi <\pi/2$ and $|\langle [X,Y] \rangle|/2=|\cos 2\phi\cos\theta|$. The inequalities can be compared directly as they saturate to the same constant value.
Let us consider when angle $\phi=0$, when two observables are orthogonal. In this case, the lower bound of (\ref{EUR}) is given by a constant $1$, whereas that of (\ref{RUR}) is determined as a function of $\theta$, $|\cos\theta|$. Then the left hand sides (LHS) of relations (\ref{DRUR}) are determined as a function of polar angle $\theta$ and azimuthal angle $\varphi$. In Fig. \ref{Graph-EUR} these functions are plotted versus $\varphi$ for fixed angles $\theta=0, 3\pi/8, 4\pi/9$ and $\theta\sim\pi/2$. It means that we take into account state vectors in a circle located half way between the north pole and the equator, depicted by the dashed blue line on Fig. \ref{Bloch}, and determined by $\theta$. This result is noteworthy. Fig. \ref{Graph-EUR} shows that the entropic uncertainty relation (EUR) (Blue) tends to move into an optimized regime as polar angle $\theta$ approaches $\pi/2$ from 0 ((a)$-$(d) in Fig. \ref{Graph-EUR}). In contrast, Robertsons uncertainty relation (RUR) (Orange) diverges. Geometrically, it can be argued that when the state vector $|\psi\rangle$ is placed in the plane of two observables (the red plane in Fig. \ref{Bloch}), the EUR in the first inequality (\ref{DRUR}) is optimised. Whereas the RUR in the second inequality of (\ref{DRUR}) is optimised when $|\psi\rangle$ is aligned along the z-axis. Especially, the state aligned along the z-axis becomes a spin coherent state whose variances of the two measurements $\sigma_x$ and $\sigma_y$ are equivalent, as a constant $=1$.
For non-orthogonal observables, i.e. $\phi\neq n \pi/2$ where $n$ is an integer, the EUR has a minimal value for a state vector lying on the x-y plane. The uncertainty relation based on the standard deviation shows divergence at $\pi/2$, since its lower bound vanishes when $|\psi\rangle$ is given by an eigenvector of observables.
Consequently, it can be said that neither of the two relations is stronger in the case of discrete observables in general. Depending upon the state provided, the EUR and the RUR characterize the trade-off relationship differently. The EUR is the optimal relation when the state is located in the same plane as the two measurements, while the RUR is optimized when the state is in the plane orthogonal to both observables. For the case of continuous variable measurements, the situation changes slightly in that the EUR for position and momentum observables is stronger than the relation based on standard deviation \cite{Biaynicki1975}.
\section{Entropic uncertainty relation for successive measurements}\label{3}
In this section, we will show the entropic uncertainty relation for successive measurements and consider the limit of our ability to measure two nondegenerate observables $X$, $Y$.
Following-on from Heisenberg's original insight, M. D. Srinivas derived the EUR for successive measurements in 2002 \cite{Srinivas2002} as follows. Consider observables $X$ and $Y$ with non-degenerate spectra,
\begin{eqnarray}\label{EURSM}
H_{\rho}(X)+H_{\mathcal{E}(\rho)}(Y) \geq -2\log c,
\end{eqnarray}
where $\mathcal{E}(\rho)=\sum_i |x_i\rangle\langle x_i|\rho|x_i\rangle\langle x_i|$. The second term $H_{\mathcal{E}(\rho)}(Y)$ is the Shannon entropy associated with the marginal of the joint probability Pr$(x_i,y_j)=\text{Tr}[ (|y_j\rangle\langle y_j|)(|x_i\rangle\langle x_i|)\rho(|x_i\rangle\langle x_i|)]$, defined by
\begin{eqnarray}
H_{\mathcal{E}(\rho)}(Y)=-\sum_j \text{Tr}[\mathcal{E}(\rho) |y_j\rangle\langle y_j|] \log \text{Tr}[\mathcal{E}(\rho) |y_j\rangle\langle y_j|]. \nonumber
\end{eqnarray}
He argued that this relation reflects statement (iii). However, it is not equivalent to Ozawa's universally valid error-disturbance relation, since it does not include the effect of the measuring process. We will propose an improved form of the EUR for successive measurements and highlight its differences with the EDUR.
Assuming that we perform a projective measurement described by nondegenerate projection operators $\{X_i\}$. In the projection postulate, an input density matrix is changed to output states determined by corresponding outcomes. The probability of obtaining an outcome $i$ is given by $\text{Pr}(x_i)=\text{Tr}[X_i \rho]$, where the input density matrix is $\rho$. In the case where we obtain the outcome $i$ after the measurement, the output state $\rho^X_i$ is given by \cite{NC10}
\begin{eqnarray}
\rho^A_i=\frac{X_i \rho X_i}{\text{Tr}[X_i \rho]},
\end{eqnarray}
according to the projection postulate.
In successive measurements, the probability of obtaining an eigenvalue $a_i$ of eigenstate $|x_i\rangle$ after the measurement of $X$ is given by $\text{Pr}(x_i)=\text{Tr}[\rho|x_i\rangle\langle x_i|]$. If we perform the measurement of $Y$ on an output state obtained just after the first measurement of $X$, we obtain an eigenvalue $y_j$ with a probability $\text{Pr}(y_j|x_i)=\text{Tr}[\rho^X_i(|y_j\rangle\langle y_j|)]$. Then the joint probability $\text{Pr}(x_i,y_j)$ of outcomes $x_i$ and $y_j$ in successive measurements is given by $\text{Pr}(x_i)\text{Pr}(y_j|x_i)$.
From the projection postulate the joint entropy of the probability distribution for the subsequent measurements is given by
\begin{eqnarray}
H_\rho(X,Y)&=&-\displaystyle\sum_{i,j}\text{Pr}(x_i)\text{Pr}(y_j|x_i) \log \text{Pr}(x_i)\text{Pr}(y_j|x_i)\nonumber.
\end{eqnarray}
The entropy $H(X,Y)$, defined in terms of the joint probability, means an amount of uncertainty is present when a state is measured by successive measurements of $X$ and $Y$. According to the sub-additivity inequality, the joint entropy has a relation with the entropy of marginal distributions of the joint probability, i.e. the LHS of the entropic uncertainty relation (\ref{EURSM}), as
\begin{eqnarray}\label{Sub-add}
H_\rho(X)+H_{\mathcal{E}(\rho)}(Y)\geq H_\rho(X,Y).
\end{eqnarray}
Furthermore, the joint entropy satisfies the following relation \cite{Srinivas2002}
\begin{eqnarray}\label{JE}
H_\rho(X,Y) \geq -2 \log c.
\end{eqnarray}
This relation implies a limitation of measuring observables $X$ and $Y$ which are not compatible with each other in successive measurements.
\begin{figure}[t]
\includegraphics[scale =0.45]{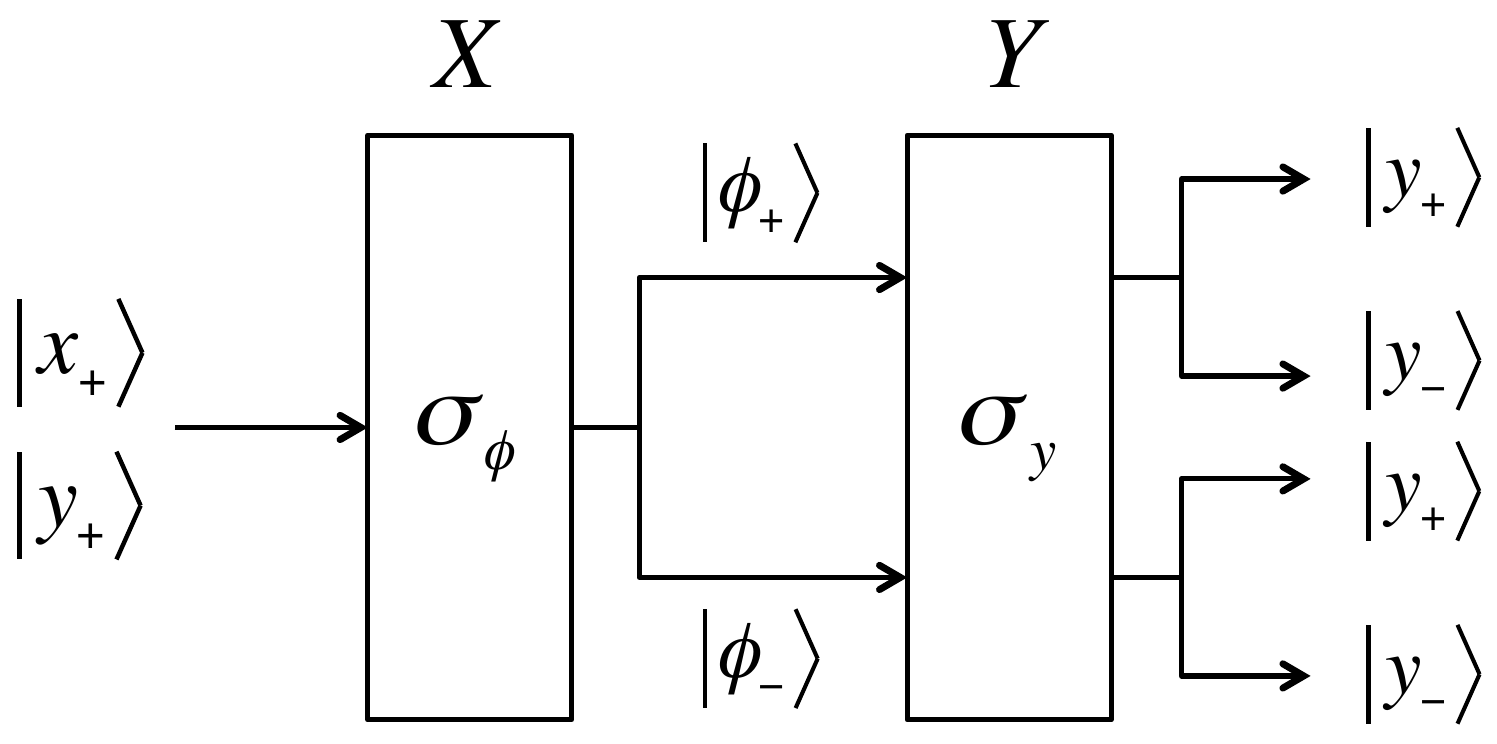}
\caption{Schematic of a successive measurement scheme for observables $X$ and $Y$, used to clarify and compare EURs ($\ref{EURSM}$), ($\ref{EURJE}$) and ($\ref{EURCE}$). After preparing the input states $|x_+\rangle$ and $|y_+\rangle$, which are eigenstates of $\sigma_x$ and $\sigma_y$ respectively, the successive measurement is assumed to measure observables $X=\sigma_\phi$ and $Y=\sigma_y$. It results in four possible outcomes.}\label{Scheme}
\end{figure}
The joint entropy, $H(X,Y)$, can be decomposed into the entropy of $X$ and the conditional entropy of $Y$ given $X$, such that
\begin{eqnarray}\label{EURJE}
H_\rho(X,Y)=H_\rho(X)+H_\rho(Y|X)\geq - 2\log c,
\end{eqnarray}
where the conditional entropy of the observable $Y$ given $X$ for a density matrix $\rho$ is defined as
\begin{eqnarray}
H_\rho(Y|X)=\sum_{i,j} \text{Pr}(x_i)\text{Pr}(y_j|x_i)\log \text{Pr}(y_j|x_i). \nonumber
\end{eqnarray}
It can be seen from relation (\ref{EURJE}) that the total uncertainty in successive measurements characterized by $H_\rho(X,Y)$ consists of the uncertainty of $X$ and the averaged uncertainty of $Y$ over outcomes $x_i$. The bound of the relation (\ref{JE}) comes from the conditional entropy, since the conditional entropy $H_\rho(Y|X)$ satisfies the following relation,
\begin{eqnarray}\label{EURCE}
H_\rho(Y|X)\geq -2 \log c
\end{eqnarray}
for nondegenerate observables $X$ and $Y$.
It also follows that it is impossible to measure incompatible observables $X$, $Y$ with certainty using successive projective measurements. Moreover the joint entropy is composed of the entropy of $X$ and the conditional entropy of $Y$ given $X$. For entropy, $H_\rho(X)$, it follows that the uncertainty characterising a density matrix $\rho$, and the conditional entropy, $H_\rho (Y|X)$, leads to an averaged uncertainty in observable $Y$ caused by the projective measurement of $X$.
\begin{figure}[t]
{\bf (a)}
\begin{minipage}[h]{1.0\linewidth}
\includegraphics[scale =0.8]{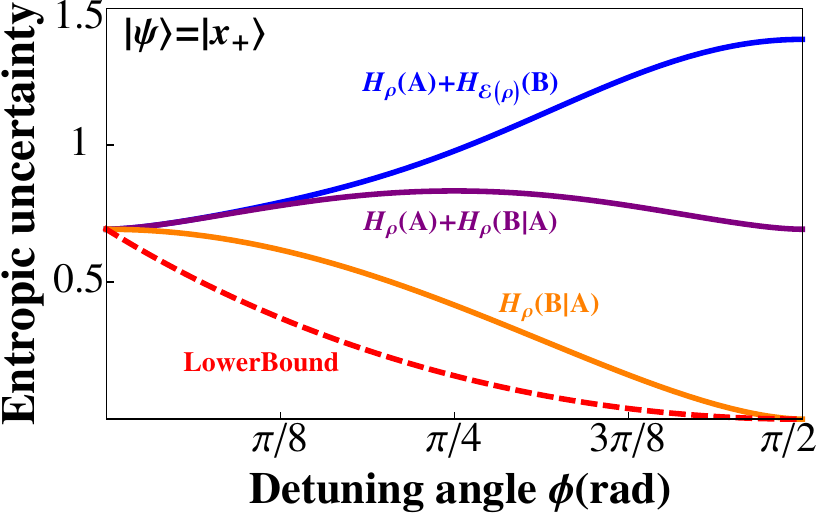}
\end{minipage}
\vspace{10 mm}
{\bf (b)}
\begin{minipage}[h]{1.0\linewidth}
\includegraphics[scale =0.8]{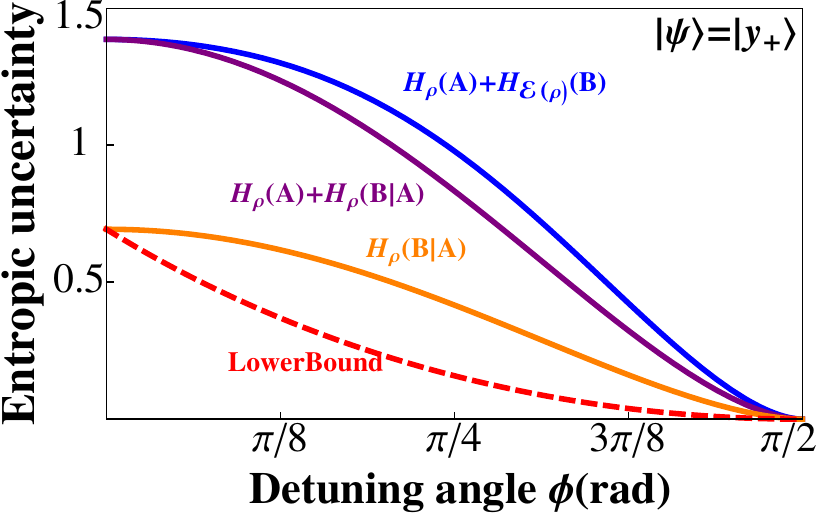}
\end{minipage}
\caption{Graphs showing the left hand side(LHS) of relations (\ref{EURSM})-(Blue) (\ref{EURJE})-(Purple), (\ref{EURCE})-(Orange) and the lower bound(Dashed red) with respect to the detuning angle $\phi$. Two input states $|x+\rangle$ and $|y+\rangle$ are considered. By inspectio
n, we find that the LHS of EUR for successive measurement (\ref{EURSM}) always has larger value than the LHS of (\ref{EURJE}) due to sub-additivity (\ref{Sub-add}). The conditional entropy has a smaller value. Three relations have zero value only when two measurements are the same at $\phi=\pi/2$, but only the conditional entropy is saturated when two observables are mutually unbiased at $\phi=0$. This means that inner products of all pairs of each eigenstate is given by $1/\sqrt{d}$, where $d$ is a dimension of the Hilbert space \cite{Wehner2010}. }\label{Graph}
\end{figure}
We will apply the above inequalities $(\ref{EURSM})$, $(\ref{EURJE})$ and $(\ref{EURCE})$ to make clear the relations among them.
Let us consider successive spin measurements assumed to satisfy the projection postulate. Measurements of $X$, $Y$ in the relations (\ref{EURSM})$\sim$(\ref{EURCE}) are designed to carry out the projective measurements of the Pauli matrices $\hat\sigma_\phi=\hat\sigma_x \cos \phi +\hat\sigma_y \sin \phi$ and $\hat\sigma_y$, respectively, as depicted in Fig. \ref{Scheme}. Since each measurement has its own eigenvectors, it projects the input state onto spin up state $|+\rangle$ or down state $|-\rangle$ after the measurements. In this way, its final result $(a_i,b_j)$ emerges among 4 possible outcomes $\{(\pm,\pm),(\pm,\mp)\}$, depicted in Fig. \ref{Scheme}.
Fig. \ref{Graph} shows the left hand side of EURs (\ref{EURSM}), (\ref{EURJE}), (\ref{EURCE}) and the calculated lower bound as a function of $\phi$. When we compare graphs of three EURs on Fig \ref{Graph}, our new relation of the conditional entropy (\ref{EURCE}) is closest to the lower bound, since the relation among them is such that
\begin{eqnarray}
H_\rho(X)+H_{\mathcal{E}(\rho)}(Y) &\geq& H_\rho(X,Y) \nonumber.\\
&=&H_\rho(X)+H_\rho(Y|X)\nonumber\\
&\geq& H_\rho(Y|X)\nonumber \\
&\geq& -2\log c\nonumber,
\end{eqnarray}
where $c=\max_{i,j}|\langle x_i|y_j\rangle|$. Three EURs have the same value when the input state is prepared in an eigenstate of the first measurement, since an outcome is determined by the corresponding eigenvalue of the input state and the first measurement does not change the input state, i.e. $H_\rho(X)=0$ and $H_{\epsilon(\rho)}(Y)=H_\rho(Y|X)$.
\section{Comparison between the EUR for successive measurements and the EDUR}\label{4}
In his proposal for the EUR for successive measurements \cite{Srinivas2002}, M. D. Srinivas said that `to explore the influence of the measurement of one observable on the uncertainties in the outcomes of another, we have to formulate an uncertainty relation for successive measurements'. However, the EUR for successive measurements does not reflect Heisenberg's microscope experiment, a thought-experiment proposed in \cite{Heisenberg1927}, since it does not consider the effect of the measuring process. On the other hand, in the error disturbance relation (\ref{UUR}), the error is defined by the distance between a POVM of an apparatus and an observable $X$, and the disturbance due to information loss in the input state caused by the measuring process \cite{Ozawa2001}. Thus, the error-disturbance relation is equivalent to statement (iii) and reflects Heisenberg's microscope experiment. This is in the sense that an effort to measure an observable $X$ exactly increases the disturbance of another observable $Y$ that is incompatible with $X$.
To formalize Heisenberg's intuition, the mean error and the disturbance are mathematically well defined in \cite{Ozawa2001} using an indirect measurement. This is because all quantum measurements can be described by an indirect measurement. In the assumption that the measuring apparatus $M$ has a family of $\{{M}_m\}$ measurement operators, the error and disturbance are defined as \cite{Ozawa2005}
\begin{eqnarray}
&&\epsilon(X)^2 =\sum_m ||{M}_m (m-X)|\psi\rangle||^2,\\
&&\eta(Y)^2 = \sum_m||[{M}_m,Y]|\psi\rangle||^2,
\end{eqnarray}
respectively, where $|\psi\rangle$ denotes an input state and $||\dots||$ denotes the norm of the state vector. Theses quantities are characterized by a measuring process realized in apparatus $M$.
\begin{figure}[t]
\begin{minipage}[h]{0.55\linewidth}
{\bf (a) $\theta=0\;\;\;\;\;\;\;\;\;\;\;\;\;\;\;\;\;\;\;\;\;$}
\end{minipage}\begin{minipage}[h]{0.25\linewidth}
{\bf (b) $\theta=\pi/4$}
\end{minipage}
\begin{minipage}[h]{0.25\linewidth}
\includegraphics[scale =0.5]{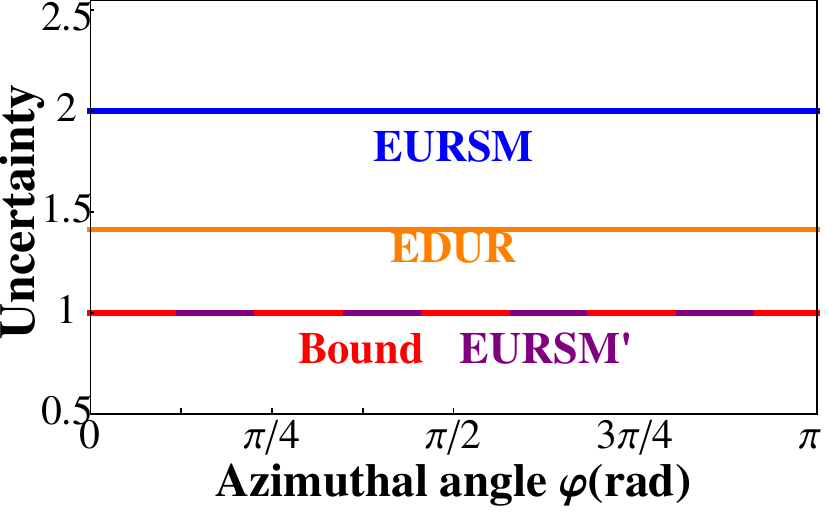}
\end{minipage}\begin{minipage}[h]{1.05\linewidth}
\includegraphics[scale =0.5]{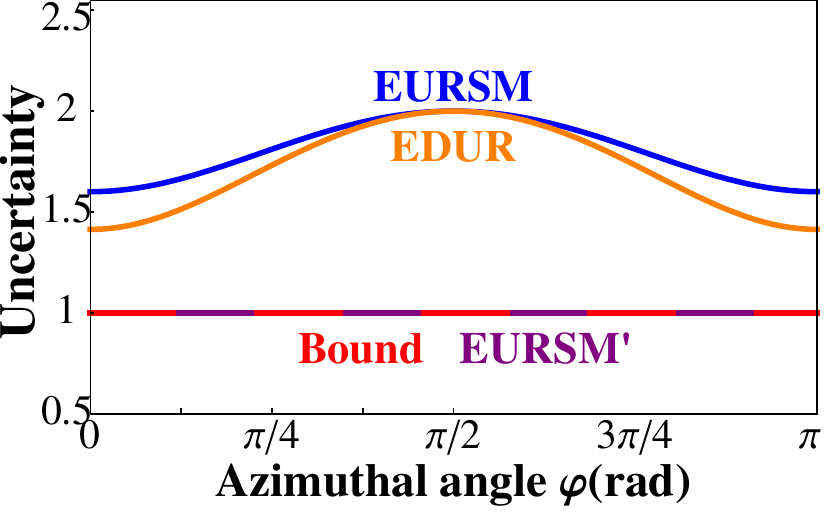}
\end{minipage}
\vspace{3 mm}
\begin{minipage}[h]{0.55\linewidth}
{\bf (c) $\theta=3\pi/8\;\;\;\;\;\;\;\;\;\;\;\;\;\;\;\;\;\;\;\;\;$}
\end{minipage}\begin{minipage}[h]{0.25\linewidth}
{\bf (d) $\theta\sim \pi/2$}
\end{minipage}
\begin{minipage}[h]{0.25\linewidth}
\includegraphics[scale =0.5]{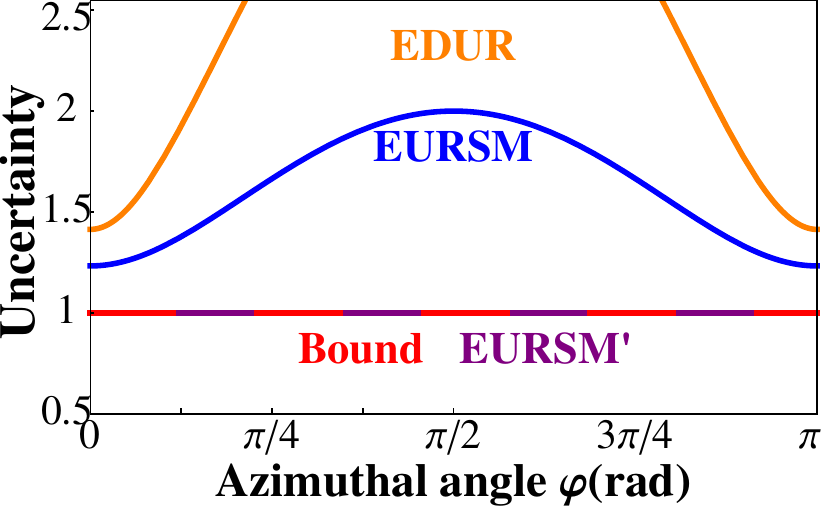}\end{minipage}\begin{minipage}[h]{1.05\linewidth}
\includegraphics[scale =0.5]{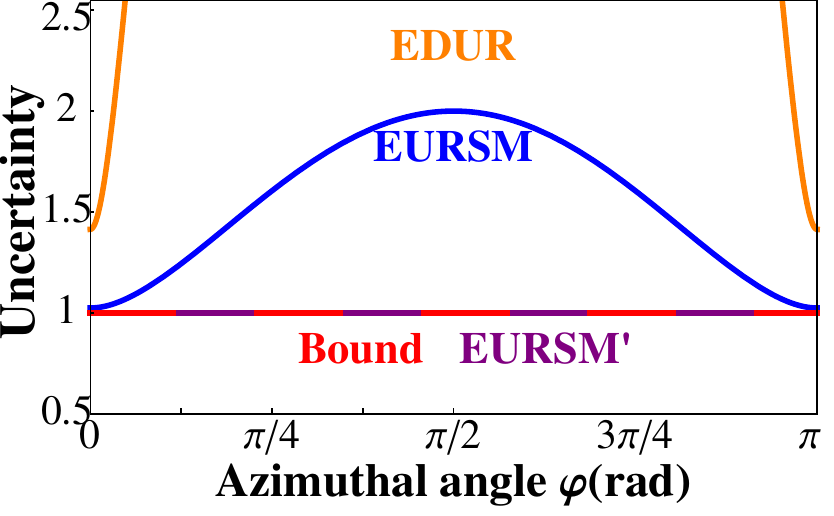}
\end{minipage}
\caption{Figures illustrating how the different EURs for successive measurements (\ref{EURJE}), (\ref{EURCE}) and the EDUR (\ref{UUR}) impose restrictions on the probabilities of outcomes of observables $X(0)$ and $Y(0)$. The LHS of relations (\ref{DEURSM}) (Blue-EURSM), (\ref{DEURSM'}) (Purple-EURSM') and (\ref{DUURSM}) (Orange-EDUR) are plotted against azimuthal angle $\varphi$, for fixed values of polar angle $\theta$ ($0$, $\pi/4$, $3\pi/8$ and $\pi/2$). By inspection, we find that (\ref{DEURSM})-EURSM decreases with respect to increasing $\theta$, whereas (\ref{DUURSM})-EDUR diverges. However, the conditional entropy of (\ref{DEURSM'})-(EURSM') is given by constant 1 for all values of $\theta$ and $\phi$. }\label{Graph-EURSM}
\end{figure}
However, from the perspective of the error-disturbance relation, the EUR for successive measurements (\ref{EURJE}) is constructed under the assumption that a measuring apparatus designed for measuring an observable $X$ precisely measures $X$. Namely, that there is no error in performing successive measurements. In this case, the EDUR (\ref{UUR}) reduces to
\begin{eqnarray}\label{UURSM}
\delta (X)\;\eta(Y)\geq\frac{1}{2}|\langle[X,Y]\rangle|,
\end{eqnarray}
since the error $\epsilon(X)$ vanishes from (\ref{UUR}).
In the assumption of precise successive measurements, the EURs (\ref{EURJE}), (\ref{EURCE}) and EDUR (\ref{UUR}) restrict possible probabilities of outcomes of $X$ and $Y$. A natural question at this stage is which of these relations places more restrictions on the probabilities. We compare them by dividing them by their own lower bounds, such that
\begin{eqnarray}
\frac{H_{|\psi\rangle}(X)+H_{|\psi\rangle}(Y|X)}{-2\log c} &\ \geq 1 \label{DEURSM} \\
\frac{H_{|\psi\rangle}(Y|X)}{-2\log c}\ \label{DEURSM'} \geq 1 \\
\frac{\delta (X)\;\eta(Y)}{|\langle[\hat{X},\hat{Y}]\rangle|/2} \geq1 , \label{DUURSM}
\end{eqnarray}
for strictly positive bounds.
Using relations (\ref{DEURSM}), (\ref{DEURSM'}) and (\ref{DUURSM}), we consider a successive measurement of observables $X(0)=\sigma_x$ and $Y(0)=\sigma_y$ with an input state vector $|\psi\rangle$. Then the probabilities of outcomes obtained by successive measurement of $X(0)$, $Y(0)$ are restricted by the uncertainty relations. In Fig. \ref{Graph-EURSM}, the LHS of relations (\ref{DUURSM}) and (\ref{DEURSM}) are plotted together against azimuthal angle $\varphi$ for fixed polar angle $\theta$. As a result, we can see in Fig. $\ref{Graph-EURSM}$ that for all $\theta$ and $\phi$, the LHS of (\ref{DEURSM'}) has the same value with the bound. This means that it imposes the highest restriction among the relations for successive measurements.
In the case of nonorthogonal observables, the EDUR divided by its lower bound has minimum value 1 at $\varphi=\phi$, and maximum value $\sqrt{1+\tan^2 \theta}$ at $\varphi=(\pi/2+\phi)$. However, the EUR for successive measurements divided by its lower bound is independent of the input state $|\psi\rangle$. Namely, it is only determined as function of $\phi$, and its value increases as $\phi$ goes to $(\pi/4+n\pi/2)$.
\section{CONCLUSIONS}\label{222}
In this work, we derive the entropic uncertainty relation for subsequent measurements and compare it with the uncertainty relations based on the standard deviation using spin measurements. A new form of EUR for successive measurements is proposed in view of Heisenberg's 1927 statement \cite{Heisenberg1927} that ``it is impossible to {\it measure position without disturbing momentum}". Much debate and effort has been expended on formalising its underlying meaning, while more recently, experiments have found different ways of demonstrating it \cite{Baek2013, Rozema2013, Sulyok2013, Ringbauer2013}. A state-independent information theoretic error-disturbance relation has also been proposed \cite{Buscemi2013}, which shows a trade-off relation between error and disturbance.
However, the EUR for successive measurements does not coincide with the EDUR and we make clear the difference between them by plotting restrictions imposed on possible probabilities of outcomes of observables in successive projective measurements without error. We can conclude that under the assumption of precise successive measurements, it is limited to obtaining outcomes with certainty. And this limitation is clarified by the relation for conditional entropy.
The realization of successive measurements was discussed in \cite{Leggett1985}. Furthermore, not all measurements fail to satisfy the projection postulate. A typical model which does not satisfy it is the photon counting measurement suggested in \cite{Ozawa2001}. In this sense, the EUR for successive measurements derived under the projection postulate cannot be applied to all measurements, although the result gives us a notion of limited measurability of successive projective measurements.
\begin{acknowledgments}
This work was supported by the National Research Foundation
of Korea (NRF) grant funded by the Korean Government
(No. NRF-2013R1A1A2010537).
\end{acknowledgments}

\bibliography{apssamp}% Produces the bibliography via BibTeX.
\end{document}